
\font\twelverm=cmr10 scaled 1200    \font\twelvei=cmmi10 scaled 1200
\font\twelvesy=cmsy10 scaled 1200   \font\twelveex=cmex10 scaled 1200
\font\twelvebf=cmbx10 scaled 1200   \font\twelvesl=cmsl10 scaled 1200
\font\twelvett=cmtt10 scaled 1200   \font\twelveit=cmti10 scaled 1200
\skewchar\twelvei='177   \skewchar\twelvesy='60
\def\twelvepoint{\normalbaselineskip=12.4pt
  \abovedisplayskip 12.4pt plus 3pt minus 9pt
  \belowdisplayskip 12.4pt plus 3pt minus 9pt
  \abovedisplayshortskip 0pt plus 3pt
  \belowdisplayshortskip 7.2pt plus 3pt minus 4pt
  \smallskipamount=3.6pt plus1.2pt minus1.2pt
  \medskipamount=7.2pt plus2.4pt minus2.4pt
  \bigskipamount=14.4pt plus4.8pt minus4.8pt
  \def\rm{\fam0\twelverm}          \def\it{\fam\itfam\twelveit}%
  \def\sl{\fam\slfam\twelvesl}     \def\bf{\fam\bffam\twelvebf}%
  \def\mit{\fam 1}                 \def\cal{\fam 2}%
  \def\tt{\twelvett}
  \textfont0=\twelverm   \scriptfont0=\tenrm   \scriptscriptfont0=\sevenrm
  \textfont1=\twelvei    \scriptfont1=\teni    \scriptscriptfont1=\seveni
  \textfont2=\twelvesy   \scriptfont2=\tensy   \scriptscriptfont2=\sevensy
  \textfont3=\twelveex   \scriptfont3=\twelveex  \scriptscriptfont3=\twelveex
  \textfont\itfam=\twelveit
  \textfont\slfam=\twelvesl
  \textfont\bffam=\twelvebf \scriptfont\bffam=\tenbf
  \scriptscriptfont\bffam=\sevenbf
  \normalbaselines\rm}

\def\beginlinemode{\endmode
  \begingroup\parskip=0pt \obeylines\def\\{\par}\def\endmode{\par\endgroup}}
\def\beginparmode{\endmode
  \begingroup \def\endmode{\par\endgroup}}
\let\endmode=\par
{\obeylines\gdef\
{}}
\def\singlespace{\baselineskip=\normalbaselineskip}
\def\oneandthreefifthsspace{\baselineskip=\normalbaselineskip
  \multiply\baselineskip by 8 \divide\baselineskip by 5}

\def\oneandahalfspace{\baselineskip=\normalbaselineskip
  \multiply\baselineskip by 3 \divide\baselineskip by 2}
\def\doublespace{\baselineskip=\normalbaselineskip \multiply\baselineskip by 2}
\newcount\firstpageno
\firstpageno=2
\footline={
\ifnum\pageno<\firstpageno{\hfil}\else{\hfil\twelverm\folio\hfil}\fi}
\let\rawfootnote=\footnote              
\def\footnote#1#2{{\rm\singlespace\parindent=0pt\rawfootnote{#1}{#2}}}
\def\raggedcenter{\leftskip=2em plus 12em \rightskip=\leftskip
  \parindent=0pt \parfillskip=0pt \spaceskip=.3333em \xspaceskip=.5em
  \pretolerance=9999 \tolerance=9999
  \hyphenpenalty=9999 \exhyphenpenalty=9999 }
\parskip=\medskipamount
\twelvepoint            
\overfullrule=0pt       
\def\preprintno#1{
 \rightline{\rm #1}}    
\def\author                     
  {\vskip 3pt plus 0.2fill \beginlinemode
   \singlespace \raggedcenter \twelvesc}
\def\affil                      
  {\vskip 3pt plus 0.1fill \beginlinemode
   \oneandahalfspace \raggedcenter \sl}
\def\abstract                   
  {\vskip 3pt plus 0.3fill \beginparmode
   \doublespace \narrower \noindent ABSTRACT: }
\def\endtitlepage               
  {\endpage                     
   \body}
\def\body                       
  {\beginparmode}               

\def\subhead#1{                 
  \vskip 0.25truein             
  {\raggedcenter #1 \par}
   \nobreak\vskip 0.1truein\nobreak}
\def\refto#1{$|{#1}$}           
\def\references                 
  {\subhead{References}         
   \beginparmode
   \frenchspacing \parindent=0pt \leftskip=1truecm
   \parskip=8pt plus 3pt \everypar{\hangindent=\parindent}}
\gdef\refis#1{\indent\hbox to 0pt{\hss#1.~}}    
\gdef\journal#1, #2, #3, 1#4#5#6{               
    {\sl #1~}{\bf #2}, #3, (1#4#5#6)}           
\def\refstylenp{                
  \gdef\refto##1{ [##1]}                                
  \gdef\refis##1{\indent\hbox to 0pt{\hss##1)~}}        
  \gdef\journal##1, ##2, ##3, ##4 {                     
     {\sl ##1~}{\bf ##2~}(##3) ##4 }}
\def\refstyleprnp{              
  \gdef\refto##1{ [##1]}                                
  \gdef\refis##1{\indent\hbox to 0pt{\hss##1)~}}        
  \gdef\journal##1, ##2, ##3, 1##4##5##6{               
    {\sl ##1~}{\bf ##2~}(1##4##5##6) ##3}}
\def\pr{\journal Phys. Rev., }

\def\prpts{\journal Phys. Rep., }
\def\np{\journal Nucl. Phys., }
\def\pl{\journal Phys. Lett., }

\def\ijmp{\journal Int. J. Mod. Phys., }
\def\endreferences{\body}
\def\endpage                    
  {\vfill\eject}
\def\endpaper                   
  {\endmode\vfill\supereject}
\def\endit
  {\endpaper\end}
\def\ref#1{Ref. #1}                     
\def\Ref#1{Ref. #1}                     
\def\begintable{\offinterlineskip\hrule}

\def\tablerule{\tablespace\noalign{\hrule}\tablespace}
\def\endtable{\hrule}

\def\m@th{\mathsurround=0pt }
\font\twelvesc=cmcsc10 scaled 1200
\def\cite#1{{#1}}
\def\(#1){(\call{#1})}
\def\call#1{{#1}}
\def\taghead#1{}
\def\leaderfill{\leaders\hbox to 1em{\hss.\hss}\hfill}
\def\twiddle{\lower.9ex\rlap{$\kern-.1em\scriptstyle\sim$}}
\def\bigtwiddle{\lower1.ex\rlap{$\sim$}}
\def\gtwid{\mathrel{\raise.3ex\hbox{$>$\kern-.75em\lower1ex\hbox{$\sim$}}}}
\def\ltwid{\mathrel{\raise.3ex\hbox{$<$\kern-.75em\lower1ex\hbox{$\sim$}}}}
\def\square{\kern1pt\vbox{\hrule height 1.2pt\hbox{\vrule width 1.2pt\hskip 3pt
   \vbox{\vskip 6pt}\hskip 3pt\vrule width 0.6pt}\hrule height 0.6pt}\kern1pt}
\catcode`@=11
\newcount\tagnumber\tagnumber=0

\immediate\newwrite\eqnfile
\newif\if@qnfile\@qnfilefalse
\def\write@qn#1{}
\def\writenew@qn#1{}
\def\w@rnwrite#1{\write@qn{#1}\message{#1}}
\def\@rrwrite#1{\write@qn{#1}\errmessage{#1}}

\def\taghead#1{\gdef\t@ghead{#1}\global\tagnumber=0}
\def\t@ghead{}

\expandafter\def\csname @qnnum-3\endcsname
  {{\t@ghead\advance\tagnumber by -3\relax\number\tagnumber}}
\expandafter\def\csname @qnnum-2\endcsname
  {{\t@ghead\advance\tagnumber by -2\relax\number\tagnumber}}
\expandafter\def\csname @qnnum-1\endcsname
  {{\t@ghead\advance\tagnumber by -1\relax\number\tagnumber}}
\expandafter\def\csname @qnnum0\endcsname
  {\t@ghead\number\tagnumber}
\expandafter\def\csname @qnnum+1\endcsname
  {{\t@ghead\advance\tagnumber by 1\relax\number\tagnumber}}
\expandafter\def\csname @qnnum+2\endcsname
  {{\t@ghead\advance\tagnumber by 2\relax\number\tagnumber}}
\expandafter\def\csname @qnnum+3\endcsname
  {{\t@ghead\advance\tagnumber by 3\relax\number\tagnumber}}

\def\equationfile{%
  \@qnfiletrue\immediate\openout\eqnfile=\jobname.eqn%
  \def\write@qn##1{\if@qnfile\immediate\write\eqnfile{##1}\fi}
  \def\writenew@qn##1{\if@qnfile\immediate\write\eqnfile
    {\noexpand\tag{##1} = (\t@ghead\number\tagnumber)}\fi}
}

\def\callall#1{\xdef#1##1{#1{\noexpand\call{##1}}}}
\def\call#1{\each@rg\callr@nge{#1}}

\def\each@rg#1#2{{\let\thecsname=#1\expandafter\first@rg#2,\end,}}
\def\first@rg#1,{\thecsname{#1}\apply@rg}
\def\apply@rg#1,{\ifx\end#1\let\next=\relax%
\else,\thecsname{#1}\let\next=\apply@rg\fi\next}

\def\callr@nge#1{\calldor@nge#1-\end-}
\def\callr@ngeat#1\end-{#1}
\def\calldor@nge#1-#2-{\ifx\end#2\@qneatspace#1 %
  \else\calll@@p{#1}{#2}\callr@ngeat\fi}
\def\calll@@p#1#2{\ifnum#1>#2{\@rrwrite{Equation range #1-#2\space is bad.}
\errhelp{If you call a series of equations by the notation M-N, then M and
N must be integers, and N must be greater than or equal to M.}}\else%
 {\count0=#1\count1=#2\advance\count1
by1\relax\expandafter\@qncall\the\count0,%
  \loop\advance\count0 by1\relax%
    \ifnum\count0<\count1,\expandafter\@qncall\the\count0,%
  \repeat}\fi}

\def\@qneatspace#1#2 {\@qncall#1#2,}
\def\@qncall#1,{\ifunc@lled{#1}{\def\next{#1}\ifx\next\empty\else
  \w@rnwrite{Equation number \noexpand\(>>#1<<) has not been defined yet.}
  >>#1<<\fi}\else\csname @qnnum#1\endcsname\fi}

\let\eqnono=\eqno
\def\eqno(#1){\tag#1}
\def\tag#1$${\eqnono(\displayt@g#1 )$$}

\def\aligntag#1\endaligntag
  $${\gdef\tag##1\\{&(##1 )\cr}\eqalignno{#1\\}$$
  \gdef\tag##1$${\eqnono(\displayt@g##1 )$$}}

\def\eqalignno#1{\displ@y \tabskip\centering
  \halign to\displaywidth{\hfil$\displaystyle{##}$\tabskip\z@skip
    &$\displaystyle{{}##}$\hfil\tabskip\centering
    &\llap{$\displayt@gpar##$}\tabskip\z@skip\crcr
    #1\crcr}}

\def\displayt@gpar(#1){(\displayt@g#1 )}

\def\displayt@g#1 {\rm\ifunc@lled{#1}\global\advance\tagnumber by1
        {\def\next{#1}\ifx\next\empty\else\expandafter
        \xdef\csname @qnnum#1\endcsname{\t@ghead\number\tagnumber}\fi}%
  \writenew@qn{#1}\t@ghead\number\tagnumber\else
        {\edef\next{\t@ghead\number\tagnumber}%
        \expandafter\ifx\csname @qnnum#1\endcsname\next\else
        \w@rnwrite{Equation \noexpand\tag{#1} is a duplicate number.}\fi}%
  \csname @qnnum#1\endcsname\fi}

\def\ifunc@lled#1{\expandafter\ifx\csname @qnnum#1\endcsname\relax}

\let\@qnend=\end\gdef\end{\if@qnfile
\immediate\write16{Equation numbers written on []\jobname.EQN.}\fi\@qnend}

\catcode`@=12
\refstyleprnp
\catcode`@=11
\newcount\r@fcount \r@fcount=0
\def\refreset{\global\r@fcount=0}
\newcount\r@fcurr
\immediate\newwrite\reffile
\newif\ifr@ffile\r@ffilefalse
\def\w@rnwrite#1{\ifr@ffile\immediate\write\reffile{#1}\fi\message{#1}}

\def\writer@f#1>>{}
\def\referencefile{
  \r@ffiletrue\immediate\openout\reffile=\jobname.ref%
  \def\writer@f##1>>{\ifr@ffile\immediate\write\reffile%
    {\noexpand\refis{##1} = \csname r@fnum##1\endcsname = %
     \expandafter\expandafter\expandafter\strip@t\expandafter%
     \meaning\csname r@ftext\csname r@fnum##1\endcsname\endcsname}\fi}%
  \def\strip@t##1>>{}}

\def\citeall#1{\xdef#1##1{#1{\noexpand\cite{##1}}}}
\def\cite#1{\each@rg\citer@nge{#1}}	

\def\each@rg#1#2{{\let\thecsname=#1\expandafter\first@rg#2,\end,}}
\def\first@rg#1,{\thecsname{#1}\apply@rg}	
\def\apply@rg#1,{\ifx\end#1\let\next=\relax
\else,\thecsname{#1}\let\next=\apply@rg\fi\next}

\def\citer@nge#1{\citedor@nge#1-\end-}	
\def\citer@ngeat#1\end-{#1}
\def\citedor@nge#1-#2-{\ifx\end#2\r@featspace#1 
  \else\citel@@p{#1}{#2}\citer@ngeat\fi}	
\def\citel@@p#1#2{\ifnum#1>#2{\errmessage{Reference range #1-#2\space is bad.}%
    \errhelp{If you cite a series of references by the notation M-N, then M and
    N must be integers, and N must be greater than or equal to M.}}\else%
 {\count0=#1\count1=#2\advance\count1
by1\relax\expandafter\r@fcite\the\count0,%
  \loop\advance\count0 by1\relax
    \ifnum\count0<\count1,\expandafter\r@fcite\the\count0,%
  \repeat}\fi}

\def\r@featspace#1#2 {\r@fcite#1#2,}	
\def\r@fcite#1,{\ifuncit@d{#1}
    \newr@f{#1}%
    \expandafter\gdef\csname r@ftext\number\r@fcount\endcsname%
                     {\message{Reference #1 to be supplied.}%
                      \writer@f#1>>#1 to be supplied.\par}%
 \fi%
 \csname r@fnum#1\endcsname}
\def\ifuncit@d#1{\expandafter\ifx\csname r@fnum#1\endcsname\relax}%
\def\newr@f#1{\global\advance\r@fcount by1%
    \expandafter\xdef\csname r@fnum#1\endcsname{\number\r@fcount}}

\let\r@fis=\refis			
\def\refis#1#2#3\par{\ifuncit@d{#1}
   \newr@f{#1}%
   \w@rnwrite{Reference #1=\number\r@fcount\space is not cited up to now.}\fi%
  \expandafter\gdef\csname r@ftext\csname r@fnum#1\endcsname\endcsname%
  {\writer@f#1>>#2#3\par}}

\def\ignoreuncited{
   \def\refis##1##2##3\par{\ifuncit@d{##1}%
     \else\expandafter\gdef\csname r@ftext\csname
r@fnum##1\endcsname\endcsname%
     {\writer@f##1>>##2##3\par}\fi}}

\def\r@ferr{\endreferences\errmessage{I was expecting to see
\noexpand\endreferences before now;  I have inserted it here.}}
\let\r@ferences=\references
\def\references{\r@ferences\def\endmode{\r@ferr\par\endgroup}}

\let\endr@ferences=\endreferences
\def\endreferences{\r@fcurr=0
  {\loop\ifnum\r@fcurr<\r@fcount
    \advance\r@fcurr by 1\relax\expandafter\r@fis\expandafter{\number\r@fcurr}%
    \csname r@ftext\number\r@fcurr\endcsname%
  \repeat}\gdef\r@ferr{}\global\r@fcount=0\endr@ferences}

\let\r@fend=\endpaper\gdef\endpaper{\ifr@ffile
\immediate\write16{Cross References written on []\jobname.REF.}\fi\r@fend}

\catcode`@=12

\citeall\refto		
\citeall\ref		%
\citeall\Ref		%

\referencefile

\def\frac#1/#2{#1 / #2}

\def\umich{Randall Physics Laboratory\\University of Michigan
\\Ann Arbor MI 48109}

\def\oneandthreefifthsspace{\baselineskip=\normalbaselineskip
  \multiply\baselineskip by 8 \divide\baselineskip by 5}

\font\titlefont=cmr10 scaled\magstep3
\def\bigtitle                      
  {\null\vskip 3pt plus 0.2fill
   \beginlinemode \doublespace \raggedcenter \titlefont}

\def\tin{{\overline t}_{I}}
%
\newbox\hdbox%
\newcount\hdrows%
\newcount\multispancount%
\newcount\ncase%
\newcount\ncols
\newcount\nrows%
\newcount\nspan%
\newcount\ntemp%
\newdimen\hdsize%
\newdimen\newhdsize%
\newdimen\parasize%
\newdimen\spreadwidth%
\newdimen\thicksize%
\newdimen\thinsize%
\newdimen\tablewidth%
\newif\ifcentertables%
\newif\ifendsize%
\newif\iffirstrow%
\newif\iftableinfo%
\newtoks\dbt%
\newtoks\hdtks%
\newtoks\savetks%
\newtoks\tableLETtokens%
\newtoks\tabletokens%
\newtoks\widthspec%
%
%
\immediate\write15{%
CP SMSG GJMSINK TEXTABLE --> TABLE MACROS V. 851121 JOB = \jobname%
}%
%
%
\tableinfotrue%
\catcode`\@=11
%
%
\def\tstrut{\vrule height3.1ex depth1.2ex width0pt}%
\def\and{\char`\&}
\def\tablerule{\noalign{\hrule height\thinsize depth0pt}}%
\thicksize=1.5pt
\thinsize=0.6pt
\def\thickrule{\noalign{\hrule height\thicksize depth0pt}}%
\def\ctr#1{\hfil\ #1\hfil}%
%
%
%
%
\tablewidth=-\maxdimen%
\spreadwidth=-\maxdimen%
\def\tabskipglue{0pt plus 1fil minus 1fil}%
%
%
\centertablestrue%
%
%
%
%
\parasize=4in%
\gdef\ARGS{########}
\gdef\headerARGS{####}
\def\@mpersand{&}
{\catcode`\|=13
\gdef\letbarzero{\let|0}
\gdef\letbartab{\def|{&&}}%
\gdef\letvbbar{\let\vb|}%
}
{\catcode`\&=4
\def\ampskip{&\omit\hfil&}
\catcode`\&=13
\let&0
\xdef\letampskip{\def&{\ampskip}}%
\gdef\letnovbamp{\let\novb&\let\tab&}
}
\def\begintable{
   \begingroup%
   \catcode`\|=13\letbartab\letvbbar%
   \catcode`\&=13\letampskip\letnovbamp%
   \def\multispan##1{
      \mscount##1%
      \multiply\mscount\tw@\advance\mscount\m@ne%
      \loop\ifnum\mscount>\@ne \sp@n\repeat%
   }
   \def\|{%
      &\omit\widevline&%
   }%
   \ruledtable
}
\long\def\ruledtable#1\endtable{%
%
%
%
   \offinterlineskip
   \tabskip 0pt
   \def\widevline{\vrule width\thicksize}
   \def\endrow{\@mpersand\omit\hfil\crnorm\@mpersand}%
   \def\crthick{\@mpersand\crnorm\thickrule\@mpersand}%
   \def\crthickneg##1{\@mpersand\crnorm\thickrule
          \noalign{{\skip0=##1\vskip-\skip0}}\@mpersand}%
   \def\crnorule{\@mpersand\crnorm\@mpersand}%
   \def\crnoruleneg##1{\@mpersand\crnorm
          \noalign{{\skip0=##1\vskip-\skip0}}\@mpersand}%
   \let\nr=\crnorule
   \def\endtable{\@mpersand\crnorm\thickrule}%
   \let\crnorm=\cr
%
%
   \edef\cr{\@mpersand\crnorm\tablerule\@mpersand}%
   \def\crneg##1{\@mpersand\crnorm\tablerule
          \noalign{{\skip0=##1\vskip-\skip0}}\@mpersand}%
   \let\ctneg=\crthickneg
   \let\nrneg=\crnoruleneg
   \the\tableLETtokens
%
%
   \tabletokens={&#1}
%
%
   \countROWS\tabletokens\into\nrows%
   \countCOLS\tabletokens\into\ncols%
%
%
   \advance\ncols by -1%
   \divide\ncols by 2%
   \advance\nrows by 1%
%
%
   \iftableinfo %
      \immediate\write16{[Nrows=\the\nrows, Ncols=\the\ncols]}%
   \fi%
%
%
   \ifcentertables
      \ifhmode \par\fi
      \line{
      \hss
   \else %
      \hbox{%
   \fi
      \vbox{%
         \makePREAMBLE{\the\ncols}
         \edef\next{\preamble}
         \let\preamble=\next
         \makeTABLE{\preamble}{\tabletokens}
      }
      \ifcentertables \hss}\else }\fi
   \endgroup
   \tablewidth=-\maxdimen
   \spreadwidth=-\maxdimen
}
\def\makeTABLE#1#2{
   {
   \let\ifmath0
   \let\header0
   \let\multispan0
%
%
   \ncase=0%
   \ifdim\tablewidth>-\maxdimen \ncase=1\fi%
   \ifdim\spreadwidth>-\maxdimen \ncase=2\fi%
   \relax
%
   \ifcase\ncase %
      \widthspec={}%
   \or %
      \widthspec=\expandafter{\expandafter t\expandafter o%
                 \the\tablewidth}%
   \else %
      \widthspec=\expandafter{\expandafter s\expandafter p\expandafter r%
                 \expandafter e\expandafter a\expandafter d%
                 \the\spreadwidth}%
   \fi %
   \xdef\next{
      \halign\the\widthspec{%
      #1
      \noalign{\hrule height\thicksize depth0pt}
      \the#2\endtable
%
      }
   }
   }
   \next
}
\def\makePREAMBLE#1{
   \ncols=#1
   \begingroup
   \let\ARGS=0
   \edef\xtp{\widevline\ARGS\tabskip\tabskipglue%
   &\ctr{\ARGS}\tstrut}
   \advance\ncols by -1
   \loop
      \ifnum\ncols>0 %
      \advance\ncols by -1%
      \edef\xtp{\xtp&\vrule width\thinsize\ARGS&\ctr{\ARGS}}%
   \repeat
   \xdef\preamble{\xtp&\widevline\ARGS\tabskip0pt%
   \crnorm}
   \endgroup
}
\def\countROWS#1\into#2{
   \let\countREGISTER=#2%
   \countREGISTER=0%
   \expandafter\ROWcount\the#1\endcount%
}%
\def\ROWcount{%
   \afterassignment\subROWcount\let\next= %
}%
\def\subROWcount{%
   \ifx\next\endcount %
      \let\next=\relax%
   \else%
      \ncase=0%
      \ifx\next\cr %
         \global\advance\countREGISTER by 1%
         \ncase=0%
      \fi%
      \ifx\next\endrow %
         \global\advance\countREGISTER by 1%
         \ncase=0%
      \fi%
      \ifx\next\crthick %
         \global\advance\countREGISTER by 1%
         \ncase=0%
      \fi%
      \ifx\next\crnorule %
         \global\advance\countREGISTER by 1%
         \ncase=0%
      \fi%
      \ifx\next\crthickneg %
         \global\advance\countREGISTER by 1%
         \ncase=0%
      \fi%
      \ifx\next\crnoruleneg %
         \global\advance\countREGISTER by 1%
         \ncase=0%
      \fi%
      \ifx\next\crneg %
         \global\advance\countREGISTER by 1%
         \ncase=0%
      \fi%
      \ifx\next\header %
         \ncase=1%
      \fi%
      \relax%
      \ifcase\ncase %
         \let\next\ROWcount%
      \or %
         \let\next\argROWskip%
      \else %
      \fi%
   \fi%
   \next%
}
\def\counthdROWS#1\into#2{%
\dvr{10}%
   \let\countREGISTER=#2%
   \countREGISTER=0%
\dvr{11}%
\dvr{13}%
   \expandafter\hdROWcount\the#1\endcount%
\dvr{12}%
}%
\def\hdROWcount{%
   \afterassignment\subhdROWcount\let\next= %
}%
\def\subhdROWcount{%
   \ifx\next\endcount %
      \let\next=\relax%
   \else%
      \ncase=0%
      \ifx\next\cr %
         \global\advance\countREGISTER by 1%
         \ncase=0%
      \fi%
      \ifx\next\endrow %
         \global\advance\countREGISTER by 1%
         \ncase=0%
      \fi%
      \ifx\next\crthick %
         \global\advance\countREGISTER by 1%
         \ncase=0%
      \fi%
      \ifx\next\crnorule %
         \global\advance\countREGISTER by 1%
         \ncase=0%
      \fi%
      \ifx\next\header %
         \ncase=1%
      \fi%
\relax%
      \ifcase\ncase %
         \let\next\hdROWcount%
      \or%
         \let\next\arghdROWskip%
      \else %
      \fi%
   \fi%
   \next%
}%
{\catcode`\|=13\letbartab
\gdef\countCOLS#1\into#2{%
   \let\countREGISTER=#2%
   \global\countREGISTER=0%
   \global\multispancount=0%
   \global\firstrowtrue
   \expandafter\COLcount\the#1\endcount%
   \global\advance\countREGISTER by 3%
   \global\advance\countREGISTER by -\multispancount
}%
\gdef\COLcount{%
   \afterassignment\subCOLcount\let\next= %
}%
{\catcode`\&=13%
\gdef\subCOLcount{%
   \ifx\next\endcount %
      \let\next=\relax%
   \else%
      \ncase=0%
      \iffirstrow
         \ifx\next& %
            \global\advance\countREGISTER by 2%
            \ncase=0%
         \fi%
         \ifx\next\span %
            \global\advance\countREGISTER by 1%
            \ncase=0%
         \fi%
         \ifx\next| %
            \global\advance\countREGISTER by 2%
            \ncase=0%
         \fi
         \ifx\next\|
            \global\advance\countREGISTER by 2%
            \ncase=0%
         \fi
         \ifx\next\multispan
            \ncase=1%
            \global\advance\multispancount by 1%
         \fi
         \ifx\next\header
            \ncase=2%
         \fi
         \ifx\next\cr       \global\firstrowfalse \fi
         \ifx\next\endrow   \global\firstrowfalse \fi
         \ifx\next\crthick  \global\firstrowfalse \fi
         \ifx\next\crnorule \global\firstrowfalse \fi
         \ifx\next\crnoruleneg \global\firstrowfalse \fi
         \ifx\next\crthickneg  \global\firstrowfalse \fi
         \ifx\next\crneg       \global\firstrowfalse \fi
      \fi
\relax
      \ifcase\ncase %
         \let\next\COLcount%
      \or %
         \let\next\spancount%
      \or %
         \let\next\argCOLskip%
      \else %
      \fi %
   \fi%
   \next%
}%
\gdef\argROWskip#1{%
   \let\next\ROWcount \next%
}
\gdef\arghdROWskip#1{%
   \let\next\ROWcount \next%
}
\gdef\argCOLskip#1{%
   \let\next\COLcount \next%
}
}
}
\def\spancount#1{
   \nspan=#1\multiply\nspan by 2\advance\nspan by -1%
   \global\advance \countREGISTER by \nspan
   \let\next\COLcount \next}%
\def\dvr#1{\relax}%
\def\header#1{%
\dvr{1}{\let\cr=\@mpersand%
\hdtks={#1}%
\counthdROWS\hdtks\into\hdrows%
\advance\hdrows by 1%
\ifnum\hdrows=0 \hdrows=1 \fi%
\dvr{5}\makehdPREAMBLE{\the\hdrows}%
\dvr{6}\getHDdimen{#1}%
{\parindent=0pt\hsize=\hdsize{\let\ifmath0%
\xdef\next{\valign{\headerpreamble #1\crnorm}}}\dvr{7}\next\dvr{8}%
}%
}\dvr{2}}
\def\makehdPREAMBLE#1{
\dvr{3}%
\hdrows=#1
{
\let\headerARGS=0%
\let\cr=\crnorm%
\edef\xtp{\vfil\hfil\hbox{\headerARGS}\hfil\vfil}%
\advance\hdrows by -1
\loop
\ifnum\hdrows>0%
\advance\hdrows by -1%
\edef\xtp{\xtp&\vfil\hfil\hbox{\headerARGS}\hfil\vfil}%
\repeat%
\xdef\headerpreamble{\xtp\crcr}%
}
\dvr{4}}
\def\getHDdimen#1{%
\hdsize=0pt%
\getsize#1\cr\end\cr%
}
\def\getsize#1\cr{%
\endsizefalse\savetks={#1}%
\expandafter\lookend\the\savetks\cr%
\relax \ifendsize \let\next\relax \else%
\setbox\hdbox=\hbox{#1}\newhdsize=1.0\wd\hdbox%
\ifdim\newhdsize>\hdsize \hdsize=\newhdsize \fi%
\let\next\getsize \fi%
\next%
}%
\def\lookend{\afterassignment\sublookend\let\looknext= }%
\def\sublookend{\relax%
\ifx\looknext\cr %
\let\looknext\relax \else %
   \relax
   \ifx\looknext\end \global\endsizetrue \fi%
   \let\looknext=\lookend%
    \fi \looknext%
}%
%
%
\def\tablelet#1{%
   \tableLETtokens=\expandafter{\the\tableLETtokens #1}%
}%
\catcode`\@=12
\def\tin{{\overline t}_I}
\def\umich{Randall Physics Laboratory,\ \ University of Michigan,\ \ Ann
Arbor MI 48109}
\def\uof{Institute for Fundamental Theory\\Department of Physics,
University of Florida\\Gainesville FL 32611}
In the minimal supersymmetric standard model, the three gauge
couplings appear to unify at a mass scale near $2 \times
10^{16}$ GeV. We investigate the possibility that intermediate
scale particle thresholds modify the running couplings so as to
increase the unification scale. By requiring consistency of
this scenario, we derive some constraints on the particle
content and locations of the intermediate thresholds. There are
remarkably few acceptable solutions with a single cleanly
defined intermediate scale far below the unification scale.
\oneandthreefifthsspace
\preprintno{UM-TH-95-01}
\preprintno{UFIFT-HEP-95-1}
\bigtitle{Raising the unification scale in supersymmetry}
\author Stephen P.~Martin
\affil\umich
\vskip.3cm
\centerline{and}
\author
Pierre Ramond
\affil\uof
\body
\abstract
In the minimal supersymmetric standard model, the three gauge
couplings appear to unify at a mass scale near $2 \times
10^{16}$ GeV. We investigate the possibility that intermediate
scale particle thresholds modify the running couplings so as to
increase the unification scale. By requiring consistency of
this scenario, we derive some constraints on the particle
content and locations of the intermediate thresholds. There are
remarkably few acceptable solutions with a single cleanly
defined intermediate scale far below the unification scale.
\endtitlepage
\oneandthreefifthsspace

\subhead{1.~Introduction}
\taghead{1.}

Data from LEP suggests that with $N=1$ supersymmetry[\cite{reviews}]
at low energy ($\sim 1$ TeV), the three gauge couplings of the
standard model converge to unify[\cite{unification}] at one scale
$M_X\approx 2 \times 10^{16}$ GeV. This apparent unification is
predicated on two assumptions. One is that the weak hypercharge
coupling is normalized to its unification into a higher rank Lie
group, such as $SU(5)$, $SO(10)$ or $E_6$. The second is the absence
of intermediate thresholds between $1$ TeV and $M_X$. This apparent
unification of couplings may  be regarded as a ``prediction" of the
low energy value of $\sin^2 \theta_W$ given the measured value of the
strong coupling constant, and is a tantalizing hint of a unifying
structure, such as superstring theory or a supersymmetric Grand
Unified Theory.

While it is clear that the three gauge couplings have a much better
chance to unify with low energy supersymmetry than without, it may be
premature to unequivocably announce their unification, and this simple
picture may have to be modified. The main reasons are the large
experimental uncertainties in the value of the QCD coupling constant
and ignorance of  the detailed structure of the supersymmetric
thresholds.

Thus it may be that the gauge couplings do not exactly unify at $M_X$.
In that case, we may want to alter this simple picture by adding  at
least one intermediate threshold between the SUSY scale and the
``unification" scale at $M_X$. The question of interest is whether the
couplings can then be made to unify at a larger scale after
introduction of the new intermediate threshold(s), caused by
particles with vector-like electroweak quantum numbers. These modify
the  running of the gauge couplings above the intermediate thresholds
to achieve true unification at the  scale $M_U$, which we take to be
larger than $M_X$. By requiring consistency of this scenario, we can
derive constraints on the particles  at the intermediate thresholds
and relations between the intermediate scales $M_X$ and $M_U$.

There are several reasons to pursue this line of inquiry. One is that
intermediate mass scales appear in many extensions of the minimal
supersymmetric standard model (MSSM), such as those which incorporate
a light invisible axion[\cite{axion}] or massive neutrinos through the
see-saw mechanism[\cite{seesaw}]. Another is to explain the near zero
values of many of the Yukawa matrix elements through  mixing the known
particles with vector-like particles. These particles may appear  at
 intermediate thresholds.

Our primary motivation, however, is superstring theory which indicates
that the unification energy should be more than one order of magnitude
above $M_X$. The effective low energy theories generated by
superstrings contain, in addition to the three chiral families, many
vector-like particles, incomplete remnants of ${\bf 27}$ and
${\bf\overline {27}}$ representations of $E_6$. These vector-like
particles have electroweak singlet masses, assumed to be, in the
absence of any special mechanism,  of the order of the highest scale
around, in this case the Planck mass. However, these theories  have a
larger invariance group than that of the MSSM, and must develop
intermediate thresholds {\it below} the string scale to break the
invariance group to that of the MSSM. This is typically achieved by
flat directions in the potential.

If the true scale of gauge coupling unification is higher than the
apparent unification scale because of intermediate scale thresholds as
assumed here, one may view the ``success" of gauge coupling
unification as  just an accident. We are implicitly taking the point
of view that it is not {\it completely} accidental, and that it is
still possible to understand gauge coupling unification through
calculable perturbative means. We therefore assume that the three
gauge couplings remain perturbative up to the unification scale $M_U$,
and that the reason behind the raising of the unification scale is not
some artifact of e.g.~stringy threshold effects, but is really due to
the presence of intermediate scale thresholds. We also assume that the
normalization of weak hypercharge is indeed the standard one
appropriate for unification with $SU(2)_L$ and $SU(3)^c$ into a simple
gauge group. (Ref. [\cite{ibanez}] explores the possibility of
different normalizations of the hypercharge as a means of raising the
unification scale.)

This paper is organized as follows. In section 2 we develop the
formalism for unification of couplings with one intermediate scale
threshold, and then for several intermediate thresholds. In section 3
we discuss the effects of various possibilities for the new particles
at the intermediate scale(s), including both new chiral superfields
and new gauge vector superfields. In section 4 we discuss the results
for one intermediate scale with  raised unification. Here we
find tight constraints on the particle content and location of the
intermediate scale. Section 5 deals with results for more than one
intermediate scale, using as an example a particular three-family
superstring model.

\subhead{2.~One-Loop Equations With New Thresholds}
\taghead{2.}

Let us begin by recalling some salient facts about the running of the
gauge couplings. Since we will be comparing the running of gauge couplings
with an intermediate scale to the ``template" case of the MSSM, it will be
sufficient to use one-loop renormalization group equations only. The three
gauge couplings run with scale according to
$$
\alpha_i^{-1}(t)=\alpha_i^{-1}(t_X)+{b_i\over 2\pi}(t-t_X)\ ,
\eqno(mssmrunning)
$$
where
$$
\alpha_i(t)={g_i^2(t)\over 4\pi}\ ,
$$
are the couplings for the three gauge groups, $i=1,2,3$ for
$U(1)_Y$, $SU(2)_L$, and $SU(3)^c$, respectively. The scale is given by
$$
t=\ln (\mu/\mu_0)\ ,
$$
where $\mu_0$ is an arbitrary reference energy, and
$$
t_X=\ln (M_X/\mu_0)\ ,
$$
is the unification scale. For $N=1$ supersymmetry we have
$$
b_i=3c_{\rm adjoint}-\sum_r c_r\ ,
\eqno(defbi)
$$
where the $c_r$'s are the Dynkin indices of the representations,
and the sum is over the left-handed chiral multiplets.
The hypercharge is normalized so that
$$
b_1=-{3\over 20}\sum_r Y_r^2\ ,
$$
corresponding to the electric charge
$$
Q=I_3+{Y\over 2}\ .
$$
For the three families and two Higgs doublets of chiral superfields in the
Minimal Supersymmetric Standard Model (MSSM), we have
$$
b_1=-{33\over 5}\ ;\qquad b_2=-1\ ;\qquad b_3=3\ .
$$

We start with the trajectories for $\alpha_1$ and $\alpha_2$ since their
values at low energies are known with the greatest accuracy. We define
$t_X$ as the scale at which these two appear to meet in the MSSM:
$$
\alpha^{-1}_X\equiv\alpha_1^{-1} (t_X) = \alpha_2^{-1} (t_X) \ .
$$
The extrapolated data, with $N=1$ supersymmetry around 1 TeV, show that
$\alpha^{-1}_X\approx 24.5$, with $M_X \approx 2 \times 10^{16}$ GeV.
However we do not assume precisely the
same value for $\alpha_3(t_X)$ at that scale, since we are assuming that the
``unification" at $M_X$ is only apparent; rather we set
$$
\alpha^{-1}_X = \alpha_3^{-1} (t_X) + \Delta \ ,
$$
introducing the parameter $\Delta $ which parameterizes our ignorance
about $\alpha_{3}(M_Z)$, our ignorance about the precise location of the
SUSY thresholds, and our negligence of two-loop effects. The present
uncertainties indicate that $$|\Delta | \le 1.5\ ,
\eqno(conservative)$$ using the most conservative estimate. We contrast
this situation by noting that without low energy supersymmetry, the same
parameters have the values $\alpha^{-1}_X\approx 42$, $M_X\approx 10^{13}$
GeV, and $\Delta \approx 5$.
\vskip .3cm
\noindent {\it Case of One Intermediate Threshold}
\vskip .3cm
Assume first only one intermediate threshold above the
supersymmetric thresholds, at the scale
$$
t_I=\ln({M_I/\mu_0})\ ;\qquad t_I<t_X\ .
$$
The previous equations are still valid as long as we are below the
intermediate threshold, that is
$$
\eqalign{ \alpha_i^{-1}(t) &
=\alpha^{-1}_X +{b_i\over 2\pi}(t-t_X)\ , \qquad\>\> (i=1,2)\cr
\alpha_3^{-1}(t) & =\alpha^{-1}_X - \Delta +{b_3\over 2\pi}(t-t_X)\ ,\cr
} \eqno(apparent)
$$
for $t\le t_I$.
At the intermediate threshold $t=t_I$,  new vector-like
particles with electroweak singlet masses at $M_I$, alter the $b_i$
coefficients to new values
$$b_i\to b_i-\delta_i\ ,\qquad i=1,2,3\ ,$$
with all $\delta_i$ positive as long as the matter is made up of chiral
superfields. We assume that their effect is to push the true unification
scale to the new value $t_U$ with $t_U > t_X$. Thus, above the
intermediate threshold, all three gauge couplings must satisfy
$$\alpha^{-1}_i(t)=\alpha^{-1}_U+{1\over
2\pi}(b_i-\delta_i)(t-t_U)\ ;\qquad t_I\le t\le t_U\ ,$$
where there is only one coupling at unification, $\alpha^{}_U$.

We have thus two ways of writing the equations for the gauge couplings
below the intermediate threshold; one is given by \(apparent), the other by
$$\alpha^{-1}_i(t)=\alpha^{-1}_U+{b_i\over 2\pi}(t-t_I)+{1\over
2\pi}(b_i-\delta_i)(t_I-t_U)\ ,\qquad i=1,2,3\ . \eqno(real) $$
Comparison of the two yields the three consistency equations
$$
\eqalign{b_i-\delta_i\left({t_U-t_I\over t_U-t_X}\right)=&
{2\pi\over t_U - t_X}\left(\alpha^{-1}_U-\alpha^{-1}_X\right)\qquad
i=1,2\ ;\cr
b_3-\delta_3\left({t_U-t_I\over t_U-t_X}\right)=&
{2\pi\over t_U - t_X }\left(\alpha^{-1}_U-\alpha^{-1}_X+\Delta \right)\
.\cr} \eqno(tresamigos)
$$
By subtracting the first two, we obtain the constraint
$${28\over 5}-(\delta_2-\delta_1)\left({t_U-t_I\over t_U-t_X}\right)=0\ ,
\eqno(constronetwo)
$$
which indicates that $\delta_2-\delta_1$ must be positive.
The difference between the second and the third equations in \(tresamigos)
yields
$$4-{2\pi\Delta \over t_U - t_X}-(\delta_3-\delta_2)\left( { t_U - t_I \over
t_U - t_X }\right)=0\ .
\eqno(constrtwothree)
$$
The remaining equation yields the value of the gauge coupling at unification
$$
\alpha^{-1}_U=\alpha^{-1}_X-{1\over 2 \pi}\left [
\delta_2 (t_U - t_I)+ t_U - t_X \right ]\ .
\eqno(alphauin)
$$
With only non-exotic matter at the intermediate threshold, the
combinations
$$q\equiv \delta_3-\delta_2
\qquad {\rm and }\qquad {2\over 5}r\equiv\delta_2-\delta_1\ ,$$
are integers.
Then \(constronetwo) and \(constrtwothree) can be rewritten as
$$
{r\over 14} = {t_U - t_X \over t_U - t_I}\ ,
\eqno(bbb)
$$
and
$$
{q\over 4} = {t_U - t_X - \pi \Delta/2 \over t_U - t_I}
\> .\eqno(aaa)
$$
It may be profitable to consider an elementary geometric derivation of
\(bbb) and \(aaa).
Consider the evolution of two inverse gauge couplings, which meet
at a scale $t_{X}$, and assume that they both change directions at a
lower scale $t_{I }$, to meet at the larger scale $t_{U}$, as shown
in Figure 1.
\vskip .5cm
\subhead{Figure 1}

The ratios of the slopes of the lines above $t_I$ satisfy, for
$\alpha_1^{-1}$ and $\alpha_2^{-1}$
$${b_2-b_1\over b^\prime_2-b^\prime_1}={OB\over
OB^\prime}={t_{X}-t_{I}\over t_{U}-t_{I}}\ ,\eqno(master)$$
from which \(bbb) follows.
We can apply the same technique to the evolution of $\alpha_2$ and $\alpha_3$
(including the near miss at $M_X$
parametrized by $\Delta $) to obtain \(aaa).

We may  think of the intermediate threshold as a ``lens" which
refocuses the lines $\alpha_1^{-1}(t)$, $\alpha_2^{-1}(t)$, and
$\alpha_3^{-1}(t)$ so that they meet at $t_U$ rather than $t_X$. If
$q> 4$, the intermediate threshold acts as a divergent lens, and the
two lines for $\alpha^{-1}_{2}$ and $\alpha^{-1}_{3}$ never intersect.
If $q=4$, the same two lines  are parallel and again never meet. Thus
we must have $q<4$ for the two curves to intersect beyond $t_I$.  In
addition, $q$ cannot be negative or $\Delta $ would be too large. This
is easy to understand, since $q<0$ corresponds to a strongly focusing
lens which would make $\alpha_2$ and $\alpha_3$ meet at a {\it lower}
scale than they would in the MSSM. To avoid having $\alpha_2$ and
$\alpha_3$ meet prematurely, $\Delta $ would have to be large and
positive when $q<0$. To see this, note that we can write
$$ \Delta  =
{1\over 2 \pi} [ 4 (t_U - t_X) - q (t_U - t_I) ]\ . \eqno(deltaa) $$
So, for instance if $q=-1$, we find that even in the case of small
hierarchies $M_X/M_I = 10$ and $M_U/M_X = 10$, one has $\Delta =2.2$,
which corresponds to a larger error than the experimental
uncertainties on $\alpha_3$ warrant. For more substantial hierarchies,
or for more negative values of $q$, the situation becomes rapidly even
worse. Thus it is sufficient to consider only the four cases,
$q=0,1,2,3$. Similarly, from \(bbb) we find that if $r\ge 14$, the
$\alpha_1^{-1}$ and $\alpha_2^{-1}$ lines will never meet, while if
$r<0$, they will meet prematurely, implying a lowered scale of
unification. If $r=0$, the unification scale is not raised and
$M_U=M_X$. Thus we have $0<r<14$.

The scale of true unification can be extracted from \(aaa) and \(bbb)
in terms of $M_X$, $M_I$, and the parameters $q, \Delta$ and $r$
respectively:
$$
M_U=M_X\left({M_X\over M_I}\right)^{q/( 4-q)}
e^{2\pi\Delta /(4-q)}\
\eqno(uniscale)
$$
$$
M_U=M_X\left({M_X\over M_I}\right)^{r/( 14-r)} \> .
\eqno(uniscaler)
$$
Taken together, these imply
$$
q = {2\over 7} r - {2\pi\Delta\over t_U - t_I}\ ,
\eqno(qrdelta)
$$
or equivalently
$$
\Delta ={1\over \pi} (t_U - t_I)\left ({r\over 7} - {q\over 2}\right )
\> .\eqno(deltaqr)
$$
If $\Delta >0$, then $r$ must be a positive integer in the
range ${7\over 2}q<r<14$. On the other hand if $\Delta $ is negative,
we have $0<r<{7q\over 2}$. The special case $\Delta =0$ yields a
non-trivial result only when $2r=7q$. In that case, for non-exotic
matter, the only solution is $q=2,r=7$, and from \(uniscale) or
\(uniscaler), $M_X$ is the geometric mean between $M_I$ and $M_U$. This
corresponds to the seemingly perverse case of the gauge couplings
unifying both with and without the intermediate threshold!
For non-zero $\Delta$,
the hierarchies of scales are summarized by the two equations
$$
\eqalignno{
{M_X\over M_I}=&\exp\left[{\pi\Delta (14-r)\over{2r-7q}}\right] \ ,
&(hierxi) \cr
{M_U\over M_X}=&\exp\left[
{\pi r\Delta  \over{2r-7q}}\right]
\ .&(hierux) \cr}
$$

Another constraint which should be taken into account is that our
equations are meaningless if the gauge couplings become too large. It is
difficult to say exactly how large is too large, but if we arbitrarily
require that $\alpha^{-1}_U\ge 2\ ,$ then given the numerical value
$\alpha^{-1}_X\approx 25$, from \(alphauin)
we obtain (safely neglecting $t_U-t_X$):
$$
\delta_2 (t_U - t_I) < 145\> .
\eqno(perturbativity)
$$

\vskip .3cm
\noindent{\it Multiple Intermediate Thresholds}
\vskip .3cm
So far we have assumed  only one intermediate threshold between 1 TeV
and $M_X$, but as previously discussed, this may not be a realistic
assumption. More generally, suppose there are $N$ distinct
intermediate mass scales $M_{Ia}$ ($a=1\ldots N$) between 1 TeV and
the unification scale. At each of these $N$ thresholds, $\delta_{1a}$,
$\delta_{2a}$, and $\delta_{3a}$ are the decreases in slope of the
running inverse gauge couplings. One may then use the master formula
\(master) iteratively to build the corresponding equations. The
results are
$$
t_U - t_X = {1\over 4} \sum_{a=1}^N q_a (t_U - t_{Ia})
+ {\pi \Delta \over 2}\ ,
$$
and
$$
t_U - t_X = {1\over 14} \sum_{a=1}^N  r_a (t_U - t_{Ia} )
\> .
$$
where $q_a = \delta_{3a} - \delta_{2a}$ and $r_a =
5(\delta_{2a} - \delta_{1a})/2$ for each of the $N$ thresholds.
Now requiring $t_U-t_X > 0$ constrains the particle content.
One may view this case as one of multiple lenses, some divergent, some
convergent.

These multiple thresholds  act like one effective
lens, which leads us to recast these equations by choosing a single
effective intermediate scale $\tin$ which should reflect the
``average" of the individual thresholds in some sense. The choice of
$\tin$ is to some extent arbitrary, as long as $\tin< t_X < t_U$, and
indeed the appropriate choice for a definition of $\tin$ depends on
the particular example being studied. Then one defines:
$$
\overline\delta_i = \sum_{a=1}^N \delta_{ia} \left (
{t_U - t_{Ia} \over t_U - \tin} \right )\ ,
\eqno(deltabars)
$$
in which each $\delta_{ia}$ is weighted more (less) when the corresponding
intermediate scale $t_{Ia}$ is lower (higher) than $\tin$.
In terms of
$$
{\overline q} \equiv \sum_{a=1}^N q_a \left (
{t_U - t_{Ia} \over t_U - \tin} \right )  \> , \qquad\>\>
{\overline r} \equiv \sum_{a=1}^N r_a \left (
{t_U - t_{Ia} \over t_U - \tin} \right )\ ,   \eqno(defqrbar)
$$
we obtain
$$
{{\overline q}\over 4} = {t_U - t_X - \pi \Delta/2 \over t_U - \tin}
\ ,\eqno(qover4)
$$
and
$$
{{\overline r} \over 14} = {t_U - t_X \over t_U - \tin}
\> .
\eqno(rover14)
$$
The gauge coupling at unification is
$$\alpha^{-1}_U=\alpha^{-1}_X-{1\over 2 \pi}\left [
\> \overline\delta_2 (t_U - \tin)+ t_U - t_X \right ]\> .
\eqno(alphaatu)
$$
Note that the above equations have the same form as in the case of a single
intermediate threshold, but with ``averaged" quantities $\tin$,
$\overline q$, $\overline r$, etc. In fact,
one still has the constraints
$$
\eqalignno{
0\le &{\overline q} < 4 \> \ ,&(conq04) \cr
0< &{\overline r} < 14  \ ,  &(conqr014) \cr
}
$$
from requiring that the coupling constants unify, but not too early.
The main difference is that ${\overline q}$,
${\overline r}$, and $\overline \delta_2$ need not be integers.
Each of the equations \(constrtwothree)-\(perturbativity)
derived in the case of a single intermediate
threshold now hold with $t_I$, $\delta_i$, $q$, $r$ replaced by
$\tin$, $\overline \delta_i$, $\overline q$, $\overline r$.

\subhead{3.~Particles at the Intermediate Threshold(s)}
\taghead{3.}

In order to analyze each case in detail, it is convenient to list the
possible representations of the new particles that generate the
intermediate thresholds, and compute their $\delta_i$ coefficients.

For the purposes of this paper, we focus on low energy theories that
could have originated from superstring theories. Thus we restrict
ourselves to representations contained in ${\bf 27}, {\bf\overline
{27}}$ and ${\bf 78}$ representations of $E_6$, under the decomposition
$$E_6\subset SU(2)_L\times SU(3)^c\times U(1)_Y\ .$$
In some string compactifications, specifically with higher level
Kac-Moody, chiral multiplets transforming as the adjoint can
survive[\cite{LYKKEN}], with their remnants appearing in the low
energy theory. The results are summarized in Table 1.
\subhead{Table 1}
\begintable
\multispan 7 \tstrut \hfill CHIRAL SUPERMULTIPLETS  \hfill
\crthick
$~{\rm
Representation}~$|$~\delta_1~$|$~\delta_2~$|$~\delta_3~$|$~q~~$
|$~r~~$|$~\#~$
\crthick
$({\bf 2},{\bf 1}^c)_{-1}+c.$|${3\over 5}$|$1$|$0$|$-1~$|$1~$|$n_1$\cr
$({\bf 1},{\bf 1}^c)_2+c.$|${6\over 5}$|$0$|$0$|$0~$|$-3~$|$n_2$\cr
$({\bf 1},{\bf \bar 3}^c)_{-{4\over 3}}+c.$|${8\over
5}$|$0$|$1$|$1~$|$-4~$|$n_3$\cr
$({\bf 1},{\bf \bar 3}^c)_{2\over 3}+c.$|${2\over
5}$|$0$|$1$|$1~$|$-1~$|$n_4$\cr
$({\bf 2},{\bf 3}^c)_{1\over 3}+c.$|${1\over 5}$|$3$|$2$|$-1~$|$7~$|$n_5$\cr
$({\bf 2},{\bf 3}^c)_{-{5\over 3}}+c.$|$5$|$3$|$2$|$-1~$|$-5~$|$n_6$\cr
$({\bf 3},{\bf 1}^c)_0$|$0$|$2$|$0$|$-2~$|$5~$|$n_7$\cr
$({\bf 1},{\bf 8}^c)_0$|$0$|$0$|$3$|$3~$|$0~$|$n_8$\endtable\vskip .3cm

The last three representations in Table 1 appear only in the
adjoint of $E_6$. We note that for all  these representations,
$5(\delta_2-\delta_1)$ is even, and $r$ is an integer.
More generally, it can be shown that all
representations for which $5(\delta_2-\delta_1)$ is odd necessarily
describe leptons with half-integer electric charges, or quarks which
yield bound states with non-integer charges. It follows from Table 1
that
$$\eqalign{q=&-n_1+n_3+n_4-n_5-n_6-2n_7+3n_8\ ,\cr
r= & \> n_1-3n_2-4n_3-n_4+7n_5-5n_6+5n_7\ ,\cr}$$
where there are $n_i$ vector-like representations at the intermediate
threshold. In the superstring compactification scenario, the vector-like
representations come from the fundamental of $E_6$.
If there are no chiral superfield remnants of the adjoint, $n_6=n_7=n_8=0$.
We see from the above that the quantity $q+r$ must be a multiple of $3$:
$$q+r=6n_5-3n_2-3n_3-6n_6+3n_7+3n_8\ .$$
We should also take into account the possibility that the gauge group
is enlarged above the intermediate scale(s). In such a case, it is possible to
identify the coupling constants for the enlarged gauge group and run the
new gauge couplings up to the high scale. However, it is not really necessary
to do so. Instead, one can simply follow the running of the three low energy
gauge couplings even though they are embedded within the larger gauge group
at high scales. Because of the assumption that the gauge couplings are
properly normalized for unification into a simple gauge group, one can take
into account the effects of gauge bosons and gauginos living at the
intermediate scales by simple step functions in the beta functions.
Therefore we generalize our
analysis to include possible vector-like remnants of a single vector
supermultiplet adjoint of $E_6$.
Table 2 is exactly the same as the first 6 rows of Table 1, except that the
entries now appear multiplied by the factor $-3$, in accordance with the
formula \(defbi) for the $b_i$, since they belong to the vector supermultiplet.
\subhead{Table 2}
\begintable \multispan 7 \tstrut \hfill VECTOR
SUPERMULTIPLETS  \hfill \crthick
$~{\rm
Representation}~$|$~\delta_1~$|$~\delta_2~$|$~\delta_3~$|$~q~$
|$~r~$|$~\#~$ \crthick
$({\bf 2},{\bf
1}^c)_{-1}+c.$|$-{9\over 5} $|$-3$|$0$|$3$|$-3$|$N_1(\le 1)$\cr
$({\bf
1},{\bf 1}^c)_2+c.$|$-{18\over 5}$|$0$|$0$|$0$|$9$|$N_2(\le 2)$\cr
$({\bf 1},{\bf \bar 3}^c)_{-{4\over 3}}+c.$|$-{24\over
5}$|$0$|$-3$|$-3$|$12$|$N_3(\le 2)$\cr
$({\bf 1},{\bf \bar
3}^c)_{2\over 3}+c.$|$-{6\over 5}$|$0$|$-3$|$-3$|$3$|$N_4(\le 1)$\cr
$({\bf 2},{\bf 3}^c)_{1\over 3}+c.$|$-{3\over 5}$|$-9$|$-6$|$3$|$-21$|$N_
5(\le 2)$\cr
$({\bf 2},{\bf 3}^c)_{-{5\over
3}}+c.$|$-15$|$-9$|$-6$|$3$|$15$|$N_6(\le 1)$
\endtable
\vskip .3cm
The numbers in parentheses reflect the multiplicity of the representation in
a single adjoint of $E_6$. The adjoint contains also five singlets with no
hypercharge, as well as the triplet which contains the $SU(2)$ gauge and
gaugino fields and the color octet of gluons and gluinos, which are
already contained in the MSSM.

In order to account for the representations already present in the
Wess-Zumino multiplets, we simply have to replace $n_i$ by
$n^\prime_i=n_i-3N_i$ Thus, the previous formulae still apply, with
the difference that the $n_i'$ can now be negative.

For each choice of possible subgroups of $E_6$ as gauge group above
$M_I$, we can write down (up to several inequivalent embeddings) the
non-zero $N_i$'s corresponding to the gauge bosons which get mass at
$M_I$. Note that we must only consider gauge groups with $N_5=N_6=0$,
because otherwise $SU(2)_L$ and $SU(3)^c$ would necessarily be unified
at $M_I$, which is in conflict with the fact that they have different
couplings at that scale. (The corresponding gauge bosons surely could
not have intermediate scale masses in any case, because of proton
decay bounds.) So, we list the  possibilities according to rank:
\vskip 1cm
\noindent {\bf Rank 4}:

Case 0: $SU(2)_L \times SU(3)^c \times U(1)$;
All $N_i = 0$.

\noindent {\bf Rank 5}:

Case 1: $SU(2)_L \times SU(3)^c \times SU(2) \times U(1)$;
(a) All $N_i = 0$ {\it or} (b) $N_2 = 1$ .

Case 2: $SU(3)_L \times SU(3)^c \times U(1)$;
$N_1 = 1$.

Case 3: $SU(2)_L \times SU(4)^c \times U(1)$;
(a) $N_3=1$
{\it or}
(b) $N_4 = 1$.

Case 4: $SU(2)_L \times SU(4)^c \times SU(2)$;
$N_3=1$, $N_2=1$.

\noindent {\bf Rank 6}:

Case 5: $SU(2)_L \times SU(3)^c \times SU(3)\times U(1)$;
$N_2=2$.

Case 6: $SU(3)_L \times SU(3)^c \times SU(2)\times U(1) $;
(a) $N_1=1$, $N_2=1$
{\it or}
(b) $N_1=1$.

Case 7: $SU(3)_L \times SU(3)^c \times SU(3) $;
$N_1=1$, $N_2=2$.

Case 8: $SU(2)_L\times SU(4)^c\times SU(2)\times U(1)$; $N_4=1$.

Case 9: $SU(2)_L \times SU(5)^c \times U(1) $;
(a) $N_2=1$, $N_3=1$, $N_4=1$
{\it or}
(b) $N_3 =2$.

Case 10: $SU(2)_L \times SU(6)^c $;
$N_2=2$, $N_3=2$, $N_4=1$

In each of cases 1, 3, 6, and 9, there are inequivalent embeddings of
the standard model gauge group, resulting in two different
possibilities for the $N_i$. There are also acceptable subgroups of
$E_6$ obtained by adding $U(1)$ factors to the rank 4 and 5
possibilities listed above. The extra $U(1)$ factors do not contribute
to the $N_i$, and do not affect the one loop renormalization group
equations for the gauge couplings.

If the gauge group above the intermediate scale is larger than the
standard model's, there must appear at the same intermediate threshold
chiral superfields containing standard model singlets, to break the
gauge group. In particular models, one must check for their presence
and that the order parameters do not produce unacceptable R-parity
violation (or baryon number violation in models which have alternative
discrete symmetries).

\subhead{4.~Results: One Intermediate Threshold}
\taghead {4.}

In the case of only one threshold, one can combine the results of the
previous two sections to enumerate the possibilities for raising the
unification scale. In section 2 we found that $q=0,1,2,3$ and that $r$
is an integer between 0 and 14, and from section 3 we found that $q+r$
is a multiple of 3. Of special interest, perhaps, are the cases for
which $M_X/M_I$ is large, so that the different scales are cleanly
separated and may be definitely associated with different physics. For
example, if $M_I$ is to be associated with an invisible axion scale,
we  expect $M_I \sim 10^{10\pm2}$ GeV, so that $M_X/M_I = 10^{6\mp
2}$. If we want the hierarchy $M_X/M_I$ to be large, without having
$|\Delta |$ be too large or giving $M_U$ outside of the correct range
between $M_X$ and the Planck scale, there are tight restrictions which
we now discuss, classified in terms of the value of $q$.

\noindent$\bullet$
When $q=0$, the unification scale does not depend on $r$. It is given by

$$ {M_U\over M_X}=\exp\left[ {\pi\Delta  \over 2}\right]  \ ,
\eqno(qzeroux) $$
while the allowed values of $r$ are multiples of
three, $r=3,6,9,12$, corresponding to

$${M_X\over
M_I}=\exp\left[{\pi\Delta \over 2}({11\over 3},{4\over 3},{5\over
9},{1\over 6})\right]\ , \eqno(qzeroxi) $$
respectively. Clearly, $\Delta $ must be positive in order to raise the
unification scale in this case, with larger values of $\Delta $
corresponding to more substantial hierarchies in $M_U/M_X$ and
$M_X/M_I$. However, note that the hierarchy $M_X/M_I$ is severely
limited unless $r=3$, and even then $M_X/M_I$ cannot exceed $6 \times
10^3$ for $\Delta  < 1.5$. In the cases $r=6,9,12$, $M_X/M_I$ cannot be large.

\noindent$\bullet$ For $q=1$, the unification scale is given by
$$
M_U={M_X^{4/3}\over M_I^{1/3}}
e^{2\pi\Delta / 3}\ ,\eqno(scaleqone)
$$
independent of $r$. The possible values of $r$ are 2, 5, 8,  and 11,
and the results for $\Delta $ in terms of the hierarchies $t_U-t_X$ and
$t_X-t_I$ from \(hierxi) and \(hierux)
are given in Table 3.
\subhead{Table 3}
\begintable
$~~r~~$|$~2~$|$~5~$|$~8~$|$~11~$\crthick
$~~\Delta /(t_U - t_X)~~$|
$~-.48~$|$~.19~$|$~.36~$|$~.43~$\cr
$~~\Delta /(t_X - t_I)~~$|
$~-.08~$|$~.11~$|$~.48~$|$~1.6~$
\endtable
\vskip .3cm
{}From Table 3 and \(scaleqone), we can see that the hierarchy $M_X/M_I$
can be very large if $r=2$ or $5$. The case $r=2$ can accomodate intermediate
scales as low as $10^8$ GeV for $\Delta $ negative, and $r=5$ can give
$M_I$ as low as $10^{10}$ GeV, for $\Delta $ positive. The case $r=8$
does not allow $M_X/M_I$ to be larger than about $20$, because otherwise
we see from Table 3 that $\Delta $ would be larger than allowed by
the experimental constraint \(conservative). The case $r=11$
does not allow $M_X/M_I$ to be large enough to be meaningful at all.

\noindent$\bullet$ For $q=2$, the unification scale is given by
$$
M_U={M_X^2\over M_I} e^{\pi\Delta }\ .
\eqno(scaleqtwo)
$$
The possible values of $r$ are 1, 4, 7, 10, and 13,
and the results for $\Delta $ are given in Table 4.
\subhead{Table 4}
\begintable
$~~r~~$|$~1~$|$~4~$|$~7~$|$~10~$|$~13~$\crthick
$~~\Delta /(t_U-t_X)~~$|
$~-3.8$|$~-.48~$|$~{0}~$|$~.19~$|$~.29~$\cr
$~~\Delta /(t_X-t_I)~~$|
$~-.29~$|$~-.19~$|$~0~$|$~.48~$|$~3.8~$
\endtable
\vskip .3cm
Clearly, in the case $r=13$ there can be no appreciable hierarchy
in $M_X/M_I$, because of the constraint \(conservative) on $\Delta $.
In the case $r=10$, the constraint on $\Delta $
implies that $M_X/M_I$ can be at most 20 or so.
The case $r=1$ can give $M_X/M_I$ as large as 200, but then does not
allow $M_U$ to be significantly larger than $M_X$. In the case $r=7$,
$\Delta $ must be zero, as we have already noted, and from \(scaleqtwo),
the hierarchy
$M_X/M_I$ must be less than $10^3$ in order that $M_U$ not exceed the Planck
scale.  The remaining case
$r=4$ can allow $M_X/M_I$ to be as large as about $3\times 10^3$, but
no larger, because otherwise we see from Table 4 that $\Delta $ would
be too negative.

\noindent$\bullet$ For $q=3$, the unification scale is given by
$$
M_U={M_X^4\over M_I^3} e^{2\pi\Delta }\ .
\eqno(scaleqthree)
$$
The possible values of $r$ are 3, 6, 9,  and 12,
and the results for $\Delta $ are given in Table 5.
\subhead{Table 5}
\begintable
$~~r~~$|$~3~$|$~6~$|$~9~$|$~12~$\crthick
$~~\Delta /(t_U-t_X)~~$|
$~-1.6$|$~-.48~$|$~-.11~$|$~.08~$\cr
$~~\Delta /(t_X-t_I)~~$|
$~-.43~$|$~-.36~$|$~-.19~$|$~.48~$
\endtable
\vskip .3cm
Clearly there is no way to get even an
order of magnitude hierarchy in $M_X/M_I$ in the case $r=12$, because otherwise
from \(scaleqthree), $M_U$ would exceed the Planck scale since $\Delta $ is
positive. The other cases have negative $\Delta $, and therefore can
accomodate a slightly larger hierarchy; for $p=3,6,9$, one can have
$M_X/M_I$ as large as 30, 70, and 50 respectively, without having
$\Delta $ be too negative or exceeding the Planck-scale bound on $M_U$.

To summarize the preceding results, there are remarkably few cases in which
one can have a large hierarchy of scales $M_X/M_I$. Only in the cases
$q=1, r=2$ and $q=1, r=5$ can one hope to
have $M_X/M_I \ge 10^4$. These appear to
be the only acceptable cases if one wishes to associate $M_I$ with
an invisible axion scale (or anything else below $10^{12}$ GeV).
The cases $q=0, r=3$ and $q=2, r=4$ and $q=2, r=7$ can give hierarchies
which are roughly in the range $M_X/M_I \sim 10^3$. All of the other
cases give smaller upper limits for $M_X/M_I$.

In the superstring scenario, an estimate of string effects
indicates that the scale of string unification should be related to the
gauge coupling through the formula [\cite{kaplunovsky}]
$$M_U\approx 2.5\sqrt{\alpha^{}_U}\times 10^{18}\> {\rm GeV}\ .
\eqno(stringscale)$$ Taking $M_X=10^{16}$ GeV, and
$\alpha_U^{-1} < \alpha^{-1}_X \approx 25$, eq.~\(stringscale)
implies that contact with the superstring can be made provided that
${M_U/M_X}>50$.

As an example, suppose we take $\Delta =0.82$ with $r=5$, $q=1$. Then eq. (4.8)
can be satisfied together with the other constraints by
$n_1=4$, $n_2=n_3=0$, $n_4=6$ and $n_5=1$. We find that
$$M_U=7.5\times 10^{17} {\rm GeV}\ ;\qquad M_I=4.4\times 10^{12} {\rm
GeV}\ ;\qquad \alpha^{-1}_U=11\ .$$
This is one of the  solutions with low $r$
for which there is only one intermediate threshold well separated from $M_X$.
It is interesting that most of the solutions with just one intermediate
scale threshold do not allow $M_X/M_I$ to be very large.

\subhead{5. Results: Several Intermediate Thresholds}
\taghead {5.}

In most superstring theories, the effective low energy gauge group at
the string scale is larger than the standard model gauge group,
and it is necessary to have several intermediate scale thresholds.
Even if there is only one order parameter associated with
the intermediate scale, the masses of the vector-like particles are related
to that order parameter by various dimensionless couplings which are certainly
not always close to unity.  This will result in some ``smearing" of the
threshold associated with each order parameter. Thus in a realistic model,
the assumption of just
one intermediate scale is probably not justified. However, we can still
profitably analyze the situation in terms of the averaged quantities
$\tin, {\overline q}, {\overline r}$, etc.~which were introduced
in section 2. These quantities summarize the effects of the intermediate scale
mass thresholds in terms of a single effective intermediate scale, with the
main difference being that ${\overline q}$ and ${\overline r}$ need not be
integers.

Let us apply our analysis to the interesting example of the 3-family
Gepner-Schimmrigk superstring model [\cite{gepschimm},\cite{wa}].
Below the string scale, the surviving gauge group is  the
$$
SU(3)_L\times SU(3)^c\times SU(3)_R
$$
subgroup of $E_6$,
corresponding to our case 7 (of section 3) with $N_1=1$ and $N_2=2$.
This gauge group is subsequently broken  to
$$
SU(2)_L \times SU(3)^c \times SU(2) \times U(1),
$$
corresponding to our case 1(b) with $N_2=1$, and then to the standard
model gauge group. There are thus at least two {\it a priori} distinct
intermediate scale order parameters associated with each reduction in rank.
The chiral superfields which survive below the string scale are
classified under the gauge group $ SU(3)_L\times SU(3)^c\times SU(3)_R $ as:
$$
\eqalign{
9 \> {\rm leptons}\qquad\sim \qquad
& ({\bf 3}, {\bf 1}, {\bf \overline 3}) \cr
6 \> {\rm mirror}\> {\rm leptons}\qquad\sim \qquad
& ({\bf \overline 3}, {\bf 1}, {\bf 3}) \cr
3 \> {\rm quarks}\qquad \sim \qquad
& ({\bf \overline 3}, {\bf 3}, {\bf 1}) \cr
3 \> {\rm antiquarks} \qquad \sim\qquad
& ({\bf 1}, {\bf\overline  3}, {\bf 3}) \cr
}
$$
and, unlike  most other string models, no mirror quarks $({\bf 3 },
{\bf \overline 3}, {\bf 1})$ or mirror antiquarks $({\bf 1 }, {\bf 3},
{\bf \overline 3})$.

This particle content includes, besides the chiral superfields for the three
families of quarks and leptons and two Higgs doublets of the minimal
supersymmetric standard model, chiral superfields corresponding to
$$
n_1 = 20; \qquad n_2 = 6;\qquad n_3 = 0; \qquad n_4 = 3; \qquad n_5 = 0\> .
$$
Combining these with the vector superfields, we have a total vector-like
particle content yielding
$$
n_1^\prime = 17; \qquad n^\prime_2 = 0;\qquad n^\prime_3 = 0;
\qquad n^\prime_4 = 3; \qquad n^\prime_5 = 0\> .
$$
Thus if all of these particles were concentrated at just
one intermediate mass scale, we would have
$$
\delta_1 = 57/5; \qquad \delta_2 = 17; \qquad \delta_3 = 3
$$
giving
$$
q_{\rm total} = -14
\qquad {\rm and} \qquad r_{\rm total} = 14
\> .\eqno(qrtotal)$$

These values lie outside the range established by \(conq04) and
\(conqr014). If the particle thresholds affect the gauge coupling
unification in a perturbative and meaningful way below $M_X$, there
must be some smearing, with the ``averaged" quantity $\overline q$
higher than $q_{\rm total}$ and $\overline r$ lower than $r_{\rm
total}$. Otherwise, from the discussion in section 2, $\alpha_3$ and
$\alpha_2$ would meet too early (just above the intermediate scale)
and $\alpha_2$ and $\alpha_1$ would {\it never} meet. It is clear that
to move things in the right direction, the contributions of the
$\delta_{2a}$ to each of $\overline q$ and $\overline r$ should be
weighted less heavily than those of $\delta_{1a}$ and $\delta_{3a}$.
This can only occur if the masses of the electrosinglet down quark
vector-like chiral superfields corresponding to $n_4$ are smaller than
the average effective scale of the other particles. (Note that in this
example, $N_3=N_4=0$.)

Let us denote by ${\overline t}_{n_1}$, ${\overline t}_{n_2}$, and
${\overline t}_{n_4}$ the arithmetic means of the scales associated with
the chiral superfields corresponding to the weak doublet vector-like
leptons, $n_1$, the weak singlet charged leptons, $n_2$, and the
down-like electroweak singlet quarks, $n_4$, respectively.
Similarly, the arithmetic means of the scales associated with the vector
supermultiplets corresponding to $N_1$ and $N_2$ are denoted by
$\overline t_{N_1}$ and $\overline t_{N_2}$.
Then it is convenient to choose for the
effective intermediate scale $\tin = {\overline t}_{n_4}$, which is just the
scale associated with the effective threshold for $\alpha_3$.
With this choice, one finds:
$$
\eqalignno{
{\overline q} &= -20 \left ( {t_U - {\overline t}_{n_1} \over t_U - \tin}
\right ) + 3
\left ( {t_U - {\overline t}_{N_1} \over t_U - \tin}
\right ) + 3    \ ,
&(stringqbar)\cr
{\overline r} &= - {\overline q} +
18 \left ( {{\overline t}_{n_2} - {\overline t}_{N_2} \over t_U - \tin}
\right )\ ,
&(stringrbar)\cr
{\overline \delta}_2 & = - {\overline q}+3\ .
&(stringdbar)\cr
}
$$

Note that $\tin$ cannot be larger than ${\overline t}_{N_1}$ or
${\overline t}_{N_2}$, because the vectorlike color triplets can only
obtain their masses at or below the scale at which the gauge group is
broken down to that of the standard model. Also, two of the vectorlike
pairs corresponding to $n_1$ must have masses at scales below
${\overline t}_{N_{1,2}}$, for the same reason. Since these contribute
negatively to the RHS of \(stringqbar), the net positive contributions
to $\overline q$ are quite limited. So we see that the only way to
obtain $0<\overline q < 4$ is for the vectorlike weak doublet leptons,
corresponding to $n_1$  to be located (on average) well above $\tin$.
{}From \(stringrbar), one can also see that the scale ${\overline
t}_{n_2}$ associated with the charged lepton chiral superfields must
also be located above ${\overline t}_{N_2}$. Finally, we see from
\(stringdbar) that if the thresholds are arranged appropriately for
gauge coupling unification, then ${\overline \delta}_2$ is
automatically not larger than 3, so that the constraint
\(perturbativity) from perturbativity of the couplings does not limit
the effective intermediate scale $\tin$ at all. Another way to see
this is to note that the slope of $\alpha_3^{-1}$ can never be
negative with this particle content. (Of course, in models with a
larger sector of strongly interacting chiral superfields, the
requirement of perturbativity can be quite important.)

If some of the chiral superfields have masses located far below $M_X$,
we have seen that some of these must include the color triplet fields
corresponding to $n_4$. This can be understood from the fact that only
these color triplets give a positive contribution to $\overline q$
among the chiral superfields of the model. One should note, however,
that there is a potential embarrassment associated with such light
color triplets; they can easily lead to proton decay at unacceptable
rates if their masses are below $M_X$, depending on their couplings to
the quark and lepton superfields of the MSSM. This can be avoided if
e.g.~one assumes the existence of a discrete symmetry[\cite{cm}]
prohibiting some or all of the baryon number and lepton number
violating couplings. Actually, the presence of vectorlike down-type
quarks below $M_X$ seems to be a fairly general
feature of string-type models in which intermediate scale thresholds
are used to raise the unification scale; see for example
[\cite{nanop},\cite{GKMR}]. One can understand this semi-quantitatively
by examining the values of $q$ and $r$ for the chiral supermultiplets
in  Table 1. Only the chiral superfields corresponding to $n_3$ and
$n_4$ can give a positive contribution to $\overline q$. However, the
superfields for $n_3$ (which are innocuous for proton decay) also give
a relatively large negative contribution to $\overline r$. Since
$\overline q$ and $\overline r$ both must be positive to raise the
unification scale, it seems that the color triplet with electric
charge $\pm 1/3$ corresponding to $n_4$ must be weighted relatively
heavily in the averaged quantities. This is another way of saying that
they are relatively light compared to the other chiral superfields
which are important in redirecting the running gauge couplings to
their new meeting point. Of course, one can always achieve a raised
unification scale fairly safely by employing only thresholds which are
close to $M_X$. In most superstring models[\cite{GKMR}], this is
almost required, since the large number of strongly interacting chiral
superfields would cause the gauge couplings to be non-perturbative if
the effective intermediate scale were much lower than about $10^{15}$
GeV.

\subhead{6.~Conclusion}

In this paper, we have examined the possibility that the true
unification scale can be raised above its apparent value of $2 \times
10^{16}$ GeV by calculable perturbative means. It might seem rather
surprising that in the MSSM the gauge couplings should appear to be
nicely headed for unification at $M_X$, only to be redirected to a new
meeting place at $M_U$. Indeed, the apparent perverseness of this
situation allows us to put some non-trivial constraints on the
scenario. In the simplest case of just one cleanly defined
intermediate scale, it is striking that the hierarchy $M_U/M_I$ is
generally quite limited. In the probably more realistic case of a
``smeared" intermediate scale or several intermediate scales, one
cannot be as precise because of the vastly increased number of unknown
parameters. However, one can still put useful constraints on the
placement of the intermediate scales and particles, by writing things
in terms of a single effective intermediate threshold. Here too, in
most realistic models based on superstrings, there is a tendency for
many of the vector-like particles to be very heavy, based simply on
the requirement that the gauge coupling remain perturbative and thus
calculable in principle at high energies.  Even in models like the one
considered in section 5, in which the absence of a large number of
vector-like strongly interacting particles causes perturbativity to be
easily maintained, one finds that it is difficult to raise the
unification scale consistently with intermediate scales much below
$M_X$. If one insists on having some chiral superfields at relatively
low intermediate scales, we find that generally these chiral
superfields include color triplets with electric charge $\pm 1/3$,
which may be dangerous for proton decay without assuming some extra
symmetry.

The difficulty in obtaining examples in which a raised unification
scale is achieved due to a relatively low intermediate scale
corresponds to our intuition that it would be surprising if the
unification of gauge couplings were totally accidental. The lower the
intermediate scale(s) are, the more we must regard the apparent
success of the unification of gauge couplings as just a perverse
accident. On the other hand, if there are intermediate scale
thresholds which are only slightly below the unification scale, then
the near perfect unification of couplings should be regarded as
partly, but certainly not completely, accidental. This scenario seems
to be the one preferred by superstring models.

The work of S.P.M. was supported in part by the United States
Department of Energy.
The work of P.R. was supported in part by the United States Department of
Energy under grant DE-FG05-86-ER40272.

\references

\refis{unification} P.~Langacker, in Proceedings of the
PASCOS90 Symposium, Eds.~P.~Nath
and S.~Reucroft, (World Scientific, Singapore 1990)
J.~Ellis, S.~Kelley, and D.~Nanopoulos, \pl 260B, 131, 1991;
U.~Amaldi, W.~de Boer, and H.~Furstenau, \pl 260B, 447, 1991;
P.~Langacker and M.~Luo, \pr D44, 817, 1991.

\refis{seesaw} M.~Gell-Mann, P.~Ramond, and R.~Slansky, in Sanibel Talk,
CALT-68-709, Feb 1979, and in {\it Supergravity},
(North Holland, Amsterdam, 1979; T.~Yanagida, in {\it Proc. of the
Workshop on
Unified Theories and Baryon Number in the Universe}, Tsukuba, Japan, 1979,
edited by A. Sawada and A. Sugamoto (KEK Report No. 79-18, Tsukuba, 1979).

\refis{axion} Jihn E. Kim, \prpts 149, 1, 1987.

\refis{GKMR} B.~R.~Greene, K.~H.~Kirklin, P.~J.~Miron, and G.~G.~Ross,
\pl B180, 69, 1986; \np B278, 667, 1986; \np B292, 606, 1987.

\refis{kaplunovsky} V.~S.~Kaplunovsky, \np B307, 145, 1988 and Erratum,
preprint Stanford-ITP-838 (1992).

\refis{gepschimm} R. Schimmrigk, \pl B193, 175, 1987; D.~Gepner,
Princeton University report, 1987.

\refis{wa} S.~Cordes and Y. Kikuchi, \ijmp A6, 5017, 1991;
J.~Wu, R.~Arnowitt and P.~Nath, \ijmp A6, 381, 1991;
J.~Wu and R.~Arnowitt, \pr D49, 4931, 1994.

\refis{nanop} J.~Ellis, C. Kounnas and D.~V.~Nanopoulos;
\np B247, 373, 1984; S.~Kelley, J.~L.~Lopez and D.~V.~Nanopoulos,
\pl B278, 140, 1992.

\refis{reviews}
For reviews, see H.~P.~Nilles,  \prpts 110, 1, 1984 and
H.~E.~Haber and G.~L.~Kane, \prpts 117, 75, 1985.

\refis{ibanez} L.~E.~Ib\'a\~nez, \pl B318, 73, 1993.

\refis{cm} D.~J.~Casta\~no and S.~P.~Martin, \pl B340, 67, 1994,
and references therein.

\refis{LYKKEN} S. Chaudhuri, S. W. Chung, and J. Lykken, FERMILAB
PUB-94-137-T, Talk given at the 2nd IFT Workshop on Yukawa Couplings
and The Origins of Mass, Gainesville, February 1994; S. Chaudhuri,
S. W. Chung, G. Hockney, and J. Lykken, Fermilab
preprint PRINT-94-0170, August 1994.

\endreferences
\endit\end